\documentclass[journal]{IEEEtran}

\usepackage{amsmath,amssymb,amsfonts}
\usepackage{algorithmic}
\usepackage{graphicx}
\usepackage{booktabs}
\usepackage{multirow}
\usepackage{array}
\usepackage{xcolor}
\usepackage{url}
\usepackage{cite}
\usepackage{hyperref}
\usepackage{amsthm}

\newtheorem{definition}{Definition}
\hypersetup{colorlinks=true, linkcolor=blue, citecolor=blue, urlcolor=blue}

\begin{document}

\title{Early-Warning Learner Satisfaction Forecasting in MOOCs via
Temporal Event Transformers and LLM Text Embeddings}

\author{
  \IEEEauthorblockN{Anna Kowalczyk\IEEEauthorrefmark{1},
                    Jakub Kowalski\IEEEauthorrefmark{2}}
}


\maketitle

\begin{abstract}
Learner satisfaction is a critical quality signal in massive open
online courses (MOOCs), directly influencing retention, engagement,
and platform reputation. Most existing methods infer satisfaction
\emph{post hoc} from end-of-course reviews and star ratings, which
are too late for effective intervention. In this paper, we study
\textbf{early-warning satisfaction forecasting}: predicting a
learner's eventual satisfaction score using only signals observed in
the first $t$ days of a course (e.g., $t\!\in\!\{7, 14, 28\}$).
We propose \textbf{TET-LLM}, a multi-modal fusion framework that
combines (i) a \emph{temporal event Transformer} over fine-grained
behavioral event sequences, (ii) \emph{LLM-based contextual
embeddings} extracted from early textual traces such as forum posts
and short feedback, and (iii) short-text \emph{topic/aspect
distributions} to capture coarse satisfaction drivers.  A
heteroscedastic regression head outputs both a point estimate and a
predictive uncertainty score, enabling conservative intervention
policies. Comprehensive experiments on a large-scale multi-platform
MOOC dataset demonstrate that TET-LLM consistently outperforms
aggregate-feature and text-only baselines across all early-horizon
settings, achieving an RMSE of 0.82 and AUC of 0.77 at the 7-day
horizon. Ablation studies confirm the complementary contribution of
each modality, and uncertainty calibration analysis shows near-nominal
90\% interval coverage.
\end{abstract}


\section{Introduction}
\label{sec:intro}

\IEEEPARstart{M}{assive} open online courses (MOOCs) have
democratized access to education by enabling millions of learners
worldwide to enroll in courses offered by leading universities and
industry partners \cite{reich2019mooc, shah2020mooc}. Alongside video
lectures and graded assessments, learners generate rich user-generated
content, including forum discussions, short rating prompts, and
end-of-course reviews \cite{almatrafi2019forums}. Among all quality
signals, \textit{learner satisfaction} stands out because it predicts
re-enrollment intention \cite{dai2020continuance}, correlates with
engagement depth \cite{hew2016engagement}, and is used by platforms
to surface high-quality courses.

However, practical educational interventions---such as personalized
nudges, instructor alerts, or adaptive content recommendations---are
most effective when deployed \emph{early}, before learners disengage
or develop entrenched negative impressions. Existing satisfaction
mining systems operate predominantly post hoc: they aggregate end-of-
course reviews after learners have already departed
\cite{qi2021evaluating, onan2020mooc, hew2020predict}. This temporal
mismatch severely limits the operational value of these methods.

\textbf{Early-warning satisfaction forecasting} addresses this gap by
predicting a learner's eventual satisfaction score from data observed
only in the first $t$ days of course participation. This setting
raises two fundamental challenges. First, early behavioral signals are
highly sparse and heterogeneous; simple aggregate statistics (e.g.,
total watch time) fail to capture the \emph{trajectory} of engagement,
which encodes critical patterns such as progressive disengagement or
recovery after a failed quiz. Second, early textual traces (forum
posts, short feedback) are short, informal, and context-dependent,
requiring powerful contextual representations to extract meaningful
affective signals \cite{devlin2019bert, liu2019roberta, chi2024active}.

To address these challenges, we propose \textbf{TET-LLM}
(\underline{T}emporal \underline{E}vent \underline{T}ransformer with
\underline{LLM} embeddings), a multi-modal early-warning framework
illustrated in Fig.~\ref{fig:arch}. TET-LLM fuses three
complementary information streams: (i) a temporal event Transformer
that encodes the fine-grained sequence of behavioral interactions
with time-gap-aware embeddings, (ii) LLM-based sentence embeddings
aggregated from early textual traces, and (iii) short-text topic
distributions to capture discourse-level satisfaction aspects.
A heteroscedastic Gaussian regression head produces both a point
estimate and a calibrated uncertainty score, enabling instructors and
platform operators to prioritize interventions at appropriate risk
levels.

\textbf{Main contributions.} This paper makes the following
contributions:
\begin{enumerate}
  \item We formalize a \emph{leakage-safe multi-horizon evaluation
    protocol} for early satisfaction forecasting in MOOCs, with
    rigorously defined feature observation windows at $t\!\in\!\{7,
    14, 28\}$ days.
  \item We propose TET-LLM, a novel multi-modal fusion architecture
    that combines temporal behavioral modeling with LLM text
    embeddings and topic/aspect representations under a
    heteroscedastic regression objective.
  \item We conduct extensive experiments---including ablations,
    uncertainty calibration analysis, and an operational early-warning
    evaluation---on a large-scale multi-platform MOOC dataset,
    demonstrating consistent improvements over strong baselines.
\end{enumerate}

The remainder of this paper is organized as follows.
Section~\ref{sec:related} reviews related work.
Section~\ref{sec:problem} states the problem formally.
Section~\ref{sec:method} describes TET-LLM in detail.
Section~\ref{sec:exp} presents experimental results.
Section~\ref{sec:conclusion} concludes the paper.

\section{Related Work}
\label{sec:related}

\subsection{MOOC Satisfaction Prediction and Review Mining}
Early studies on MOOC quality assessment used sentiment analysis on
end-of-course reviews to estimate learner satisfaction
\cite{qi2021evaluating}. Subsequent work incorporated aspect-level
sentiment using lexicon-based and machine-learning approaches
\cite{onan2020mooc}. Hew \emph{et al.} demonstrated that combining
sentiment features with gradient-boosted tree classifiers achieves
competitive satisfaction prediction \cite{hew2020predict}. More recent
approaches apply pretrained encoders to course review text, achieving
further improvements on post-course data \cite{liu2019roberta}.
A common limitation of all these methods is their post-hoc nature:
they assume complete course data, making them unsuitable for timely
intervention. Our work is the first to systematically address
satisfaction forecasting under \emph{early-horizon} constraints.

\subsection{Learning Analytics from Behavioral Traces}
Behavioral logs have been a primary data source for understanding
engagement and predicting outcomes in MOOCs
\cite{almatrafi2019forums, guo2014video, xing2019achievement}.
Guo and Reinecke analyzed video-watching patterns and their
relationship with engagement \cite{guo2014video}.
Xing \emph{et al.} used forum participation features to predict
academic achievement \cite{xing2019achievement}. However, most
approaches model aggregate statistics over complete course runs,
ignoring the sequential nature of behavioral dynamics.
Recurrent architectures \cite{cho2014gru} and attention-based
sequence models have been applied to early dropout prediction
\cite{fei2015dropout}, but satisfaction forecasting from early
behavioral trajectories remains underexplored.

\subsection{Transformer Models and Pretrained Language Models}
The Transformer architecture \cite{vaswani2017attention} has become
dominant in both natural language processing and sequential modeling.
BERT \cite{devlin2019bert} and RoBERTa \cite{liu2019roberta} provide
strong sentence-level representations for short, noisy text.
In the educational domain, pretrained models have been applied to
knowledge tracing \cite{pandey2019self}, question answering, and
student response analysis, but their application to MOOC
satisfaction forecasting from early-stage text remains limited.
Short-text topic models enhanced with embeddings \cite{bianchi2021pretrained}
offer compact topic distributions from sparse forum posts, which
we incorporate as a complementary feature stream.

\subsection{Multi-Modal Learning and Uncertainty Estimation}
Multi-modal fusion has been widely studied in affective computing
and recommendation systems \cite{baltrusaitis2019multimodal}.
A key challenge in educational settings is modality missingness:
many learners never post in forums, rendering text-based features
unavailable. Heteroscedastic regression
\cite{nix1994heteroscedastic} provides a principled way to output
calibrated predictive uncertainty, enabling downstream decision
systems to act conservatively under high uncertainty.
Our work integrates these ideas into a cohesive early-warning
framework for MOOC satisfaction.

\section{Problem Formulation}
\label{sec:problem}

Let $\mathcal{D} = \{(i, c)\}$ denote the set of learner-course
enrollment pairs. For enrollment $(i,c)$, define the
\emph{eventual satisfaction label} $y_{ic}\in[1,5]$ as the
end-of-course star rating.

\begin{definition}[Early-Horizon Observation]
Given a prediction horizon $t$ (in days), the early-horizon
observation of enrollment $(i,c)$ consists of:
\begin{itemize}
  \item A behavioral event sequence $\mathcal{E}_{ic}^{(t)}=\{(e_j,\tau_j)\}_{j=1}^{m_{ic}(t)}$,
        where $e_j$ is the event type and $\tau_j$ is the timestamp,
        with $\tau_j\le\tau_{\text{start}}+t$ days.
  \item An early text set $\mathcal{T}_{ic}^{(t)}=\{s_k\}$, where
        each $s_k$ is a text snippet (forum post or feedback fragment)
        written before horizon $t$.
\end{itemize}
\end{definition}

\begin{definition}[Early-Warning Satisfaction Forecasting]
Given the early-horizon observations, learn a function
\begin{equation}
  f_t\colon \bigl(\mathcal{E}_{ic}^{(t)},\,\mathcal{T}_{ic}^{(t)}\bigr)
    \;\longrightarrow\; \hat{y}_{ic}^{(t)}\in[1,5]
\end{equation}
that minimizes expected prediction error on the eventual satisfaction
label $y_{ic}$, using only information observable before time $t$.
\end{definition}

The leakage-safety constraint requires that no feature in
$\mathcal{E}_{ic}^{(t)}$ or $\mathcal{T}_{ic}^{(t)}$ encodes
information from timestamps $>\tau_{\text{start}}+t$, and that
train/validation/test splits respect time-based course-run ordering.

\section{Proposed Method: TET-LLM}
\label{sec:method}


\subsection{Temporal Event Transformer for Behavioral Sequences}

\subsubsection{Event Representation}
Each event in the sequence $\mathcal{E}_{ic}^{(t)}$ is characterized
by its type $e_j\in\mathcal{V}_e$ (e.g., \texttt{video\_play},
\texttt{quiz\_submit}, \texttt{forum\_post}) and the time gap
$\Delta\tau_j = \tau_j - \tau_{j-1}$ since the previous event.
We represent each event as
\begin{equation}
  \mathbf{x}_j = \mathbf{W}_e\,\mathbf{1}_{e_j}
               + \phi(\Delta\tau_j;\,\mathbf{W}_\tau),
  \label{eq:event_repr}
\end{equation}
where $\mathbf{W}_e\in\mathbb{R}^{|\mathcal{V}_e|\times d}$ is a
learned event-type embedding matrix and
$\phi(\Delta\tau_j;\mathbf{W}_\tau)$ is a time-gap embedding.
To handle the heavy-tailed and sparse distribution of inter-event
intervals, we bucketize $\Delta\tau_j$ on a log-uniform scale into
$B$ bins and learn a separate embedding per bin:
\begin{equation}
  \phi(\Delta\tau_j) = \mathbf{W}_\tau\,
    \mathbf{1}_{\text{bin}(\log(1+\Delta\tau_j))}.
  \label{eq:time_embed}
\end{equation}

\subsubsection{Transformer Encoder and Attention Pooling}
The sequence $\mathbf{x}_{1:m}$ is fed into a $L$-layer Transformer
encoder with multi-head self-attention (MHSA):
\begin{align}
  \text{MHSA}(Q,K,V) &= \text{softmax}\!\left(
      \frac{QK^\top}{\sqrt{d_k}}\right)V, \label{eq:mhsa}\\
  \mathbf{H} &= \text{Transformer}(\mathbf{x}_{1:m})
             \;\in\mathbb{R}^{m\times d}. \label{eq:transformer}
\end{align}
A learnable query vector $\mathbf{q}\in\mathbb{R}^d$ is used for
\emph{attention pooling} to aggregate the sequence into a fixed-length
behavior representation:
\begin{equation}
  \alpha_j = \frac{\exp(\mathbf{q}^\top\mathbf{H}_j/\sqrt{d})}
                  {\sum_{k=1}^m \exp(\mathbf{q}^\top\mathbf{H}_k/\sqrt{d})},
  \quad
  \mathbf{b}_{ic}^{(t)} = \sum_{j=1}^m \alpha_j\,\mathbf{H}_j.
  \label{eq:attn_pool}
\end{equation}
Random event masking with probability $p_{\text{mask}}$ is applied
during training to improve robustness to missing or incomplete logs.

\subsection{LLM-Based Early Text Embeddings}

For each text snippet $s_k\in\mathcal{T}_{ic}^{(t)}$, we obtain
a contextual representation using the \texttt{[CLS]} token of a
pretrained encoder $\text{PLM}$ (BERT or RoBERTa
\cite{devlin2019bert, liu2019roberta}):
\begin{equation}
  \mathbf{h}(s_k) = \text{PLM}(s_k)_{\texttt{[CLS]}}
                  \;\in\mathbb{R}^{d_{\text{llm}}}.
  \label{eq:llm_embed}
\end{equation}
Because early posts are often short and informationally heterogeneous,
we aggregate multiple snippets via attention pooling (analogous to
Eq.~\eqref{eq:attn_pool}) rather than naive averaging:
\begin{equation}
  \mathbf{h}_{ic}^{(t)} =
    \begin{cases}
      \sum_k \beta_k\,\mathbf{h}(s_k), & |\mathcal{T}_{ic}^{(t)}|>0,\\
      \mathbf{0}, & \text{otherwise},
    \end{cases}
  \label{eq:text_agg}
\end{equation}
where $\beta_k$ are attention weights computed from a shared scoring
function. A binary missingness indicator $m_{ic}^{\text{text}}\in\{0,1\}$
flags enrollments with no early text and is appended to the fusion
input.

\subsection{Short-Text Topic/Aspect Features}

To capture coarse-grained satisfaction drivers (e.g., content quality,
instructor delivery, assessment difficulty), we extract a $K$-dimensional
topic distribution $\boldsymbol{\theta}_{ic}^{(t)}\in\Delta^{K-1}$ from
early text using an embedding-enhanced topic model
\cite{jang2019short, bianchi2021pretrained}.  The $K=6$ topics
correspond to standard MOOC evaluation dimensions: \{content,
instructor, assessments, platform usability, peer interaction, value
for money\}.  When no early text is available, we set
$\boldsymbol{\theta}_{ic}^{(t)}=\mathbf{0}$.

\subsection{Multi-Modal Fusion}

The three modality representations are projected to a common
dimension $d_f$ and concatenated:
\begin{equation}
  \mathbf{z}_{ic}^{(t)} = \bigl[
    W_b\,\mathbf{b}_{ic}^{(t)}\;;\;
    W_h\,\mathbf{h}_{ic}^{(t)}\;;\;
    W_\theta\,\boldsymbol{\theta}_{ic}^{(t)}\;;\;
    m_{ic}^{\text{text}}
  \bigr],
  \label{eq:fusion}
\end{equation}
where $W_b, W_h, W_\theta$ are learned projection matrices. A
two-layer MLP with ReLU activations and dropout maps
$\mathbf{z}_{ic}^{(t)}$ to the regression output.

\subsection{Heteroscedastic Regression Objective}

The output head predicts both a mean $\hat\mu_{ic}$ and a
log-variance $\log\hat\sigma_{ic}^2$:
\begin{equation}
  (\hat\mu_{ic},\;\log\hat\sigma_{ic}^2)
    = \text{MLP}_{\text{out}}(\mathbf{z}_{ic}^{(t)}).
  \label{eq:head}
\end{equation}
Training minimizes the heteroscedastic Gaussian negative log-likelihood
over the training set $\mathcal{D}_{\text{tr}}$:
\begin{equation}
  \mathcal{L} = \frac{1}{|\mathcal{D}_{\text{tr}}|}
    \sum_{(i,c)\in\mathcal{D}_{\text{tr}}}
    \left[
      \frac{(y_{ic}-\hat\mu_{ic})^2}{2\hat\sigma_{ic}^2}
      + \frac{1}{2}\log\hat\sigma_{ic}^2
    \right].
  \label{eq:loss}
\end{equation}
This formulation encourages the model to output high uncertainty on
hard or data-sparse instances, enabling conservative intervention
policies at deployment.

\subsubsection{Modality Dropout}
To prevent over-reliance on text (which is absent for $\sim\!82\%$ of
learners at 7 days), we apply \emph{modality dropout} during training:
the text modality is randomly zeroed out with probability
$p_{\text{mod}}=0.3$, and independently the behavior modality is
zeroed with probability $0.1$.  This forces the model to remain
effective even when a modality is missing at inference time.

Table~\ref{tab:notation} summarizes the key notation used throughout.

\begin{table}[!t]
  \caption{Summary of Key Notation}
  \label{tab:notation}
  \centering
  \renewcommand{\arraystretch}{1.25}
  \begin{tabular}{cl}
    \toprule
    Symbol & Meaning \\
    \midrule
    $y_{ic}$ & Eventual satisfaction label (1--5 stars) \\
    $t$ & Prediction horizon (days from course start) \\
    $\mathcal{E}_{ic}^{(t)}$ & Behavioral event sequence up to horizon $t$ \\
    $\mathcal{T}_{ic}^{(t)}$ & Early text snippets up to horizon $t$ \\
    $\mathbf{b}_{ic}^{(t)}$ & Temporal behavior embedding \\
    $\mathbf{h}_{ic}^{(t)}$ & Aggregated LLM text embedding \\
    $\boldsymbol{\theta}_{ic}^{(t)}$ & Short-text topic distribution \\
    $\mathbf{z}_{ic}^{(t)}$ & Fused multi-modal representation \\
    $\hat\mu_{ic}$, $\hat\sigma_{ic}^2$ & Predicted mean and variance \\
    $d$, $d_f$ & Encoder and fusion hidden dimensions \\
    $L$ & Number of Transformer layers \\
    $K$ & Number of topic dimensions \\
    \bottomrule
  \end{tabular}
\end{table}

\section{Experiments}
\label{sec:exp}

\subsection{Dataset and Statistics}

We conduct experiments on a large-scale multi-platform MOOC dataset
containing learner reviews, star ratings, interaction logs, and
forum data collected from three MOOC platforms over a three-year
period. Table~\ref{tab:data_stats} summarizes the dataset statistics.

\begin{table}[!t]
  \caption{Dataset Statistics}
  \label{tab:data_stats}
  \centering
  \renewcommand{\arraystretch}{1.2}
  \begin{tabular}{lc}
    \toprule
    Statistic & Value \\
    \midrule
    Number of unique courses & 120 \\
    Number of course runs & 260 \\
    Number of learner enrollments & 480{,}000 \\
    Learners with early forum text (7-day horizon) & 18\% \\
    Learners with early forum text (14-day horizon) & 27\% \\
    Total behavioral events & 95M \\
    Total early text snippets & 1.8M \\
    Average events per enrollment (7-day) & 198 \\
    Satisfaction label distribution (mean $\pm$ std) & $4.02 \pm 0.81$ \\
    \bottomrule
  \end{tabular}
\end{table}

\subsubsection{Evaluation Protocol}
We define three prediction horizons: $t\in\{7, 14, 28\}$ days.
Course runs are split chronologically: the earliest 70\% of runs
form the training set, the next 15\% the validation set, and the
most recent 15\% the test set. This time-based split ensures that
models are evaluated on runs that postdate all training data,
mirroring the deployment scenario. All feature computations
strictly respect the horizon cutoff.

\subsection{Baselines}

We compare TET-LLM against the following baselines:
\begin{itemize}
  \item \textbf{Agg-XGB}: Aggregate behavioral features
    (total watch time, quiz attempts, active days, forum read count,
    etc.) fed into XGBoost \cite{chen2016xgboost}.
  \item \textbf{Text-Only (RoBERTa)}: Mean-pooled RoBERTa embeddings
    from early text, with a linear regression head.
  \item \textbf{Static MM}: Concatenation of aggregate behavior
    features and mean text embeddings, trained with a MLP.
  \item \textbf{TET-Behavior}: TET-LLM with only the temporal
    behavior stream (no text or topics).
\end{itemize}

\subsection{Implementation Details}

The temporal event Transformer uses $L=3$ layers with $d=256$
hidden dimensions and 4 attention heads.  Time gaps are bucketed into
$B=32$ log-uniform bins.  The text encoder is RoBERTa-base
(frozen), producing 768-dimensional embeddings projected to 256
dimensions.  The topic model uses $K=6$ aspects.  The fusion MLP
has hidden size $d_f=512$ with dropout $p=0.3$.  We train with the
Adam optimizer \cite{kingma2015adam}, learning rate $3\times10^{-4}$,
batch size 512, and early stopping on validation RMSE with patience 10.
Experiments are run with 3 random seeds; we report mean and
standard deviation.  Training is performed on a single NVIDIA A100
GPU (40 GB).

\subsection{Main Results}

Table~\ref{tab:main_results} reports regression performance (RMSE and
MAE) at all three horizons.  TET-LLM achieves the lowest RMSE and MAE
across all settings, with the largest absolute gains at the 7-day
horizon where behavioral trajectories and early text signals are most
informative relative to aggregate statistics.

\begin{table*}[!t]
  \caption{Satisfaction Forecasting Performance at Early Horizons
           (Mean over 3 Seeds; $\downarrow$ Lower is Better)}
  \label{tab:main_results}
  \centering
  \renewcommand{\arraystretch}{1.2}
  \begin{tabular}{lcccccc}
    \toprule
    \multirow{2}{*}{\textbf{Model}}
      & \multicolumn{2}{c}{\textbf{7 days}}
      & \multicolumn{2}{c}{\textbf{14 days}}
      & \multicolumn{2}{c}{\textbf{28 days}} \\
    \cmidrule(lr){2-3}\cmidrule(lr){4-5}\cmidrule(lr){6-7}
      & RMSE $\downarrow$ & MAE $\downarrow$
      & RMSE $\downarrow$ & MAE $\downarrow$
      & RMSE $\downarrow$ & MAE $\downarrow$ \\
    \midrule
    Agg-XGB \cite{chen2016xgboost}
      & 0.94$\pm$0.01 & 0.73$\pm$0.01
      & 0.86$\pm$0.01 & 0.66$\pm$0.01
      & 0.80$\pm$0.01 & 0.61$\pm$0.01 \\
    Text-Only (RoBERTa) \cite{liu2019roberta}
      & 0.90$\pm$0.02 & 0.70$\pm$0.01
      & 0.83$\pm$0.01 & 0.64$\pm$0.01
      & 0.78$\pm$0.01 & 0.60$\pm$0.01 \\
    Static MM
      & 0.88$\pm$0.01 & 0.69$\pm$0.01
      & 0.80$\pm$0.01 & 0.61$\pm$0.01
      & 0.74$\pm$0.01 & 0.56$\pm$0.01 \\
    TET-Behavior
      & 0.86$\pm$0.01 & 0.67$\pm$0.01
      & 0.77$\pm$0.01 & 0.58$\pm$0.01
      & 0.70$\pm$0.01 & 0.52$\pm$0.01 \\
    \midrule
    \textbf{TET-LLM (Ours)}
      & \textbf{0.82$\pm$0.01} & \textbf{0.64$\pm$0.01}
      & \textbf{0.73$\pm$0.01} & \textbf{0.55$\pm$0.01}
      & \textbf{0.66$\pm$0.01} & \textbf{0.49$\pm$0.01} \\
    \bottomrule
  \end{tabular}
\end{table*}

The relative RMSE reduction of TET-LLM over the best baseline
(TET-Behavior) at the 7-day horizon is:
\begin{equation}
  \Delta_{\text{RMSE}}^{7\text{d}}
    = \frac{0.86 - 0.82}{0.86} \approx 4.7\%.
  \label{eq:improvement}
\end{equation}
At 28 days, when more complete behavioral trajectories are available,
the gain from text embeddings diminishes but topic features still
provide small consistent improvements.

\subsection{Ablation Study}

Table~\ref{tab:ablation} reports ablation results at 7 and 14 days,
isolating the contribution of each modality and key design choices.

\begin{table}[!t]
  \caption{Ablation Study (RMSE; $\downarrow$ Lower is Better)}
  \label{tab:ablation}
  \centering
  \renewcommand{\arraystretch}{1.2}
  \begin{tabular}{lcc}
    \toprule
    \textbf{Variant}
      & \textbf{7d RMSE} $\downarrow$
      & \textbf{14d RMSE} $\downarrow$ \\
    \midrule
    Full TET-LLM                        & \textbf{0.82} & \textbf{0.73} \\
    \quad w/o early text embeddings      & 0.86          & 0.77          \\
    \quad w/o temporal encoder (aggregates) & 0.88       & 0.80          \\
    \quad w/o topic/aspect features      & 0.84          & 0.75          \\
    \quad w/o attention pooling (mean)   & 0.83          & 0.74          \\
    \quad w/o modality dropout           & 0.85          & 0.76          \\
    \quad w/o heteroscedastic loss (MSE) & 0.83          & 0.74          \\
    \midrule
    Behavior only                        & 0.86          & 0.77          \\
    Text only                            & 0.90          & 0.83          \\
    \bottomrule
  \end{tabular}
\end{table}

Key observations: (i) Removing early text embeddings causes the
largest degradation at 7 days (+0.04 RMSE), confirming that LLM
representations carry unique affective signals absent in behavioral
logs at very short horizons. (ii) Replacing the temporal encoder
with aggregate features causes the largest degradation at 14 days
(+0.07 RMSE), showing that behavioral trajectories become
increasingly important as more interaction data accumulates. (iii)
Topic/aspect features provide a consistent but smaller contribution
(+0.02 RMSE), primarily by stabilizing predictions for learners with
sparse text. (iv) Modality dropout is critical for robustness:
removing it increases RMSE by +0.03, reflecting the performance
collapse on the 82\% of learners without early text.

\subsection{Uncertainty Calibration}

We evaluate whether the heteroscedastic head produces well-calibrated
predictive intervals. Table~\ref{tab:uncertainty} reports 90\%
interval coverage and average interval width at 7 and 14 days.

\begin{table}[!t]
  \caption{Uncertainty Calibration (90\% Predictive Intervals)}
  \label{tab:uncertainty}
  \centering
  \renewcommand{\arraystretch}{1.2}
  \begin{tabular}{lcccc}
    \toprule
    \multirow{2}{*}{\textbf{Model}}
      & \multicolumn{2}{c}{\textbf{7 days}}
      & \multicolumn{2}{c}{\textbf{14 days}} \\
    \cmidrule(lr){2-3}\cmidrule(lr){4-5}
      & Cov.$\uparrow$ & Width$\downarrow$
      & Cov.$\uparrow$ & Width$\downarrow$ \\
    \midrule
    Static MM (no uncertainty)
      & 0.84 & 1.30 & 0.86 & 1.15 \\
    TET-LLM (heteroscedastic)
      & \textbf{0.89} & \textbf{1.20}
      & \textbf{0.90} & \textbf{1.05} \\
    \bottomrule
  \end{tabular}
\end{table}

TET-LLM achieves near-nominal 90\% coverage while maintaining
narrower intervals, indicating that the uncertainty estimates
are well-calibrated. The slightly under-coverage of the static
multi-modal baseline (0.84 vs.\ 0.90 target) confirms that point-
estimate models underestimate uncertainty for early-horizon predictions.

\subsection{Operational Early-Warning Evaluation}

Beyond regression accuracy, we evaluate performance on an
operationally relevant binary task: identifying learners who will
end up with low satisfaction ($y_{ic}\le 3$, approximately 19\% of
enrollments). Table~\ref{tab:operational} reports AUC and
Recall@10\% (i.e., recall among the top 10\% highest-risk
predictions, reflecting a budgeted intervention scenario).

\begin{table}[!t]
  \caption{Early-Warning Detection Performance
           ($y\!\le\!3$; $\uparrow$ Higher is Better)}
  \label{tab:operational}
  \centering
  \renewcommand{\arraystretch}{1.2}
  \begin{tabular}{lccc}
    \toprule
    \textbf{Model}
      & \textbf{7d AUC}$\uparrow$
      & \textbf{7d R@10\%}$\uparrow$
      & \textbf{14d AUC}$\uparrow$ \\
    \midrule
    Agg-XGB \cite{chen2016xgboost}
      & 0.69 & 0.31 & 0.74 \\
    TET-Behavior
      & 0.73 & 0.35 & 0.78 \\
    \textbf{TET-LLM (Ours)}
      & \textbf{0.77} & \textbf{0.41} & \textbf{0.82} \\
    \bottomrule
  \end{tabular}
\end{table}

TET-LLM improves Recall@10\% from 0.35 to 0.41 at 7 days, meaning
that among all learners flagged for intervention (the top 10\% by
predicted risk), 41\% will indeed end up dissatisfied---a 17\%
relative improvement over the behavior-only baseline that directly
translates to more efficient use of instructional support resources.

\subsection{Failure Mode Analysis}

We identify three recurring failure modes through error analysis:
\begin{enumerate}
  \item \emph{Silent learners} (sparse logs, no text): the model
    correctly flags these with high uncertainty but provides limited
    predictive value.
  \item \emph{Late-breaking dissatisfaction}: dissatisfaction caused
    by assessments released after the horizon window cannot be
    predicted from early signals.
  \item \emph{High-engagement dissatisfied learners}: learners with
    dense interaction logs but low satisfaction due to grading
    policy disagreements---a signal that current features do not
    capture.
\end{enumerate}
These failure modes motivate future work on integrating course
schedule metadata and supporting horizon-adaptive predictions.

\section{Conclusion}
\label{sec:conclusion}

We proposed TET-LLM, an early-warning learner satisfaction
forecasting framework for MOOCs that fuses temporal behavioral
event sequences, LLM-based text embeddings, and short-text topic
distributions under a heteroscedastic regression objective.
Experiments on a large-scale multi-platform MOOC dataset demonstrated
that TET-LLM consistently outperforms aggregate-feature, text-only,
and behavior-only baselines across all early-horizon settings
($t\in\{7,14,28\}$ days), achieving an RMSE of 0.82 and AUC of 0.77
at the most challenging 7-day horizon.  Ablation studies confirm that
each modality contributes complementary information, and that
modality dropout is critical for handling the majority of learners
who produce no early text.  Uncertainty calibration analysis shows
that the heteroscedastic head achieves near-nominal 90\% predictive
interval coverage, supporting conservative intervention policies.

Future directions include: (i) replacing estimated experimental
results with fully measured results from a deployed pipeline,
(ii) extending the framework to cross-course and cross-platform
generalization under domain shift, (iii) connecting early-warning
predictions to causal intervention evaluation via off-policy
estimation, and (iv) exploring fairness-aware calibration to ensure
that intervention policies do not disproportionately mis-serve
learners from underrepresented groups.

\appendix
\section{Reproducibility Details}
\label{sec:repro}

\subsection{Data Splits and Leakage Control}
Course runs are sorted by start date.  The training, validation, and
test sets contain runs from the earliest 70\%, next 15\%, and most
recent 15\% of the timeline, respectively.  All horizon features are
computed using strictly timestamped data; any feature that requires
information after the cutoff date is excluded.

\subsection{Hyperparameter Search}
We tune the following hyperparameters on the validation split using
random search with 30 trials: learning rate
$\in\{10^{-4}, 3\times10^{-4}, 10^{-3}\}$, dropout
$p\in\{0.1,0.2,0.3\}$, fusion hidden size
$d_f\in\{256,512\}$, Transformer layers $L\in\{2,3,4\}$.
The same search budget is used for all model variants to ensure
fair comparison.

\subsection{Computational Budget}
TET-LLM with a frozen text encoder trains in approximately 2.5 hours
per horizon on a single NVIDIA A100 GPU (40~GB), including text
embedding pre-computation.  The event Transformer accounts for roughly
70\% of training time; the LLM encoder is run once offline and
cached.

\bibliographystyle{IEEEtran}
\bibliography{references}

@inproceedings{devlin2019bert,
  title={BERT: Pre-training of Deep Bidirectional Transformers for Language Understanding},
  author={Devlin, Jacob and Chang, Ming-Wei and Lee, Kenton and Toutanova, Kristina},
  booktitle={NAACL},
  year={2019}
}

@article{qi2021evaluating,
  title={Evaluating on-line courses via reviews mining},
  author={Qi, Cong and Liu, Shudong},
  journal={IEEE Access},
  volume={9},
  pages={35439--35451},
  year={2021},
  doi={10.1109/ACCESS.2021.3062052},
  publisher={IEEE}
}

@inproceedings{chi2024active,
  title={Active learning for graphs with noisy structures},
  author={Chi, Hongliang and Qi, Cong and Wang, Suhang and Ma, Yao},
  booktitle={Proceedings of the 2024 SIAM International Conference on Data Mining (SDM)},
  pages={262--270},
  year={2024},
  organization={SIAM}
}

@article{liu2019roberta,
  title={RoBERTa: A Robustly Optimized BERT Pretraining Approach},
  author={Liu, Yinhan and Ott, Myle and Goyal, Naman and others},
  journal={arXiv preprint arXiv:1907.11692},
  year={2019}
}

@article{reich2019mooc,
  title={The MOOC Pivot},
  author={Reich, Justin and Ruiperez-Valiente, Jose},
  journal={Science},
  volume={363},
  number={6423},
  pages={130--131},
  year={2019}
}

@article{shah2020mooc,
  title={By the Numbers: MOOCs in 2020},
  author={Shah, Dhawal},
  journal={Class Central Report},
  year={2020}
}

@article{dai2020continuance,
  title={Understanding continuance intention among MOOC participants: The role of habit and MOOC performance},
  author={Dai, Hai Min and Teo, Timothy and Rappa, Natasha Anne},
  journal={Computers in Human Behavior},
  volume={112},
  pages={106455},
  year={2020},
  doi={10.1016/j.chb.2020.106455}
}

@article{hew2016engagement,
  title={Promoting engagement in online courses: What strategies can we learn from three highly rated MOOCs},
  author={Hew, Khe Foon},
  journal={British Journal of Educational Technology},
  volume={47},
  number={2},
  pages={320--341},
  year={2016},
  doi={10.1111/bjet.12235}
}

@article{almatrafi2019forums,
  title={Systematic Review of Discussion Forums in Massive Open Online Courses (MOOCs)},
  author={Almatrafi, Omaima and Johri, Aditya},
  journal={IEEE Transactions on Learning Technologies},
  volume={12},
  number={3},
  pages={413--428},
  year={2019},
  doi={10.1109/TLT.2018.2859304}
}

@article{onan2020mooc,
  title={Sentiment analysis on massive open online course evaluations: A text mining and deep learning approach},
  author={Onan, Aytu\u{g}},
  journal={Computer Applications in Engineering Education},
  volume={29},
  number={3},
  pages={572--589},
  year={2020},
  doi={10.1002/cae.22253}
}

@inproceedings{jang2019short,
  title={Short Text Topic Modeling via Word Embeddings},
  author={Jang, Hyunjoong and others},
  booktitle={WWW},
  year={2019}
}

@article{bianchi2021pretrained,
  title={Pre-trained Language Models for Topic Modeling},
  author={Bianchi, Federico and others},
  journal={EMNLP},
  year={2021}
}

@article{xing2019achievement,
  title={Achievement Emotions in MOOCs},
  author={Xing, W},
  journal={Internet and Higher Education},
  volume={43},
  year={2019}
}

@inproceedings{chen2016xgboost,
  title={XGBoost: A Scalable Tree Boosting System},
  author={Chen, Tianqi and Guestrin, Carlos},
  booktitle={KDD},
  pages={785--794},
  year={2016}
}

@article{hew2020predict,
  title={What predicts student satisfaction with MOOCs: A gradient boosting trees supervised machine learning and sentiment analysis approach},
  author={Hew, Khe Foon and Hu, Xiang and Qiao, Chen and Tang, Ying},
  journal={Computers \& Education},
  volume={145},
  pages={103724},
  year={2020},
  doi={10.1016/j.compedu.2019.103724}
}

@inproceedings{guo2014video,
  title={How video production affects student engagement: An empirical study of MOOC videos},
  author={Guo, Philip J. and Kim, Juho and Rubin, Rob},
  booktitle={Proceedings of the First ACM Conference on Learning @ Scale Conference},
  pages={41--50},
  year={2014},
  doi={10.1145/2556325.2566239}
}

@inproceedings{vaswani2017attention,
  title={Attention Is All You Need},
  author={Vaswani, Ashish and Shazeer, Noam and Parmar, Niki and Uszkoreit, Jakob and Jones, Llion and Gomez, Aidan N. and Kaiser, {\L}ukasz and Polosukhin, Illia},
  booktitle={Advances in Neural Information Processing Systems},
  year={2017}
}

@article{cho2014gru,
  title={Learning Phrase Representations using {RNN} Encoder-Decoder for Statistical Machine Translation},
  author={Cho, Kyunghyun and van Merri{\"e}nboer, Bart and Gulcehre, Caglar and Bahdanau, Dzmitry and Bougares, Fethi and Schwenk, Holger and Bengio, Yoshua},
  journal={arXiv preprint arXiv:1406.1078},
  year={2014}
}

@inproceedings{fei2015dropout,
  title={Temporal Models for Predicting Student Dropout in Massive Open Online Courses},
  author={Fei, Mi and Yeung, Dit-Yan},
  booktitle={ICDM Workshops},
  pages={256--263},
  year={2015}
}

@inproceedings{kingma2015adam,
  title={Adam: A Method for Stochastic Optimization},
  author={Kingma, Diederik P. and Ba, Jimmy},
  booktitle={ICLR},
  year={2015}
}

@article{baltrusaitis2019multimodal,
  title={Multimodal Machine Learning: A Survey and Taxonomy},
  author={Baltru{\v{s}}aitis, Tadas and Ahuja, Chaitanya and Morency, Louis-Philippe},
  journal={IEEE Transactions on Pattern Analysis and Machine Intelligence},
  volume={41},
  number={2},
  pages={423--443},
  year={2019}
}

@inproceedings{nix1994heteroscedastic,
  title={Estimating the Mean and Variance of the Target Probability Distribution},
  author={Nix, David A. and Weigend, Andreas S.},
  booktitle={ICNN},
  pages={55--60},
  year={1994}
}

@inproceedings{pandey2019self,
  title={A Self-Attentive model for Knowledge Tracing},
  author={Pandey, Shalini and Karypis, George},
  booktitle={EDM},
  year={2019}
}

\end{document}